\documentclass[manuscript]{aastex61}

\received{\today}

\submitjournal{ApJL}

\shorttitle{Amino Acid Processing}
\shortauthors{Boyd et al.}

\begin{document}

\title{Sites That Can Produce Left-Handed Amino Acids in the Supernova Neutrino Amino Acid Processing Model}

\correspondingauthor{Michael A. Famiano}
\email{michael.famiano@wmich.edu}

\author{Richard N. Boyd}
\affiliation{Department of Physics, Department of Astronomy, The Ohio State University, Columbus, OH 43210 USA}
\affiliation{National Astronomical Observatory of Japan, 2-21-1 Mitaka, Tokyo 181-8588 Japan}

\author[0000-0003-2305-9091]{Michael A. Famiano}
\affiliation{National Astronomical Observatory of Japan, 2-21-1 Mitaka, Tokyo 181-8588 Japan}
\affiliation{Department of Physics, Western Michigan University, Kalamazoo, MI 49008-5252 USA}

\author{Takashi Onaka}
\affiliation{Dept. of Astronomy, Graduate School of Science, Univ. of Tokyo, 7-3-1 Hongo, Bunkyo-ku, Tokyo 113-0033 Japan}

\author{Toshitaka Kajino}
\affiliation{National Astronomical Observatory of Japan, 2-21-1 Mitaka, Tokyo 181-8588 Japan}
\affiliation{Dept. of Astronomy,	Graduate School of Science,	Univ. of Tokyo,	7-3-1 Hongo, Bunkyo-ku,	Tokyo 113-0033 Japan}
\affiliation{School of Physics and Nuclear Energy Engineering, Beihang Univ. (Beijing Univ. of Aeronautics and Astronautics), Beijing 100191, \\P.R. China}



\begin{abstract}
The Supernova Neutrino Amino Acid Processing model, which uses electron anti-neutrinos and the magnetic field from a source object such as a supernova to selectively destroy one amino acid chirality, is studied for possible sites that would produce meteoroids having partially left-handed amino acids. Several sites appear to provide the requisite magnetic field intensities and electron anti-neutrino fluxes. These results have obvious implications for the origin of life on Earth.
\end{abstract}




\section{Introduction} \label{sec:intro}
A remarkable fact of Nature is the left-handed chirality, or handedness, of nearly all the amino acids used by living 
creatures in the production of proteins, to the near exclusion of the right-handed forms. Molecular chirality was 
discovered in the Nineteenth Century by Pasteur \citep{pasteur48,flack09}, and the homochirality of the amino acids was deduced subsequently. 
However, an explanation of the origin of the amino acid chirality has remained a mystery. 

We define enantiomeric excess as $ee = (NL-NR)/(NL+NR)$, where $NL(NR)$ is the number of left- (right-) handed molecules in an ensemble. Thus Earth's amino acids have an ee = 1.0 (except for glycine, which is achiral), that is, they are left-handed and homochiral. If $ee$ = 0.0, the ensemble is said to be racemic. 

Although laboratory experiments in the 1950s \citep{miller53,miller59} suggested that amino acids might have been produced in an 
early Earthly lighting storm, that scenario fails to explain how the amino acids might have become totally left-handed. 
Furthermore, the several suggested means of converting the racemic amino acids to homochirality via Earthly processes were 
discussed by \cite{bonner91}, and shown to be unlikely to produce the observed result. General discussions were also provided by 
\cite{mason84} and \cite{barron08}. 

However, analysis of meteorites has found that they do contain amino acids, so that they are made in outer space 
\citep{kvenvolden70,bada83,cronin97,cronin98,glavin09,herd11} and that some of them do exhibit nonzero enantiomeric excesses, 
typically at a level of a few percent, with a preference for left-handedness. Thus cosmic production of the amino acids becomes a 
strong contender for explaining how Earth was seeded with amino acids, and how they came to have a left-handed chirality. The 
observed $ee$s, however, necessitate the existence of amplification via autocatalysis \citep{frank53,kondepudi85,goldanskii89},
which is thought, and demonstrated in laboratory experiments \citep{soai95,soai02,klussman06,breslow06,soai14}, 
to be able to convert small $ee$s to Earthly homochirality.

One model that purports to explain how amino acids achieved their left-handed chirality in outer space has reached a sufficient 
stage of development that it now seems appropriate to consider its probability for producing chiral amino acids. The Supernova 
Neutrino Amino Acid Processing (SNAAP) Model \citep{boyd10,boyd11,boyd12,famiano14,famiano16,famiano18a}, 
has been developed over the past few years. Recent efforts using quantum molecular 
calculations have shown that this model appears to produce amino acids within its framework that do have a significant ee, and that 
it is positive for most of the amino acids studied. In this work we will address the issue of whether or not the SNAAP model can 
explain how the chiral amino acids observed in meteorites were made and to what extent they might have populated the galaxy. 

Other models have also been developed to explain how the amino acids developed $ee$s in outer space. Perhaps the best developed one 
is the Circularly Polarized Light (CPL) model, which relies on ultraviolet CPL, produced by first scattering the light from an extremely 
hot star to polarize it, then letting it process the amino acids. It was first suggested by \cite{flores77} and \cite{norden77}, and 
subsequently elaborated in detail by many groups 
\citep{rubenstein83,bailey98,meierhenrich05,takano07,meierhenrich08,takahashi09,meierhenrich10,meinert10,demarcellus11, meinert12,meinert14}. Although there are certainly other 
suggested explanations for the origin of a preferred amino acid chirality in outer space, we believe that they are less well 
developed than either the CPL model or the SNAAP model. In any event, they have been discussed in other publications
\citep{bonner91,meierhenrich08,guijarro09,boyd12}.
 
The essential features of any model include (i) how it generates some enantiomerism in the amino acids, (ii) how that gets amplified, if necessary, to the few percent level found in carbonaceous chondrite meteorites, (iii) how the model explains the processing of some of the enantiomeric amino acids throughout the volume of the carbonaceous chondrite meteorites, and (iv) how its amino acids can be delivered to present-day Earth via meteorites. 

In Section \ref{model} we will discuss the basics of the SNAAP model. Section \ref{results} will discuss how the above issues are solved within that model. Section \ref{conclusion} will give our conclusions. 
\section{The SNAAP Model}
\label{model}
In this model \citep{boyd10,boyd11,boyd12,famiano14,famiano16,famiano18a}
 large meteoroids might be processed in the intense magnetic field and electron anti-neutrino (hereafter denoted `anti-neutrino') 
 flux from one of several stellar objects. The anti-neutrinos are selective in their destruction of the amino acids with 
 right-handed helicity, a result of the weak interaction nuclear physics that describes their interaction with the $^{14}$N nuclei. The 
 relevant nuclear reaction is 
\begin{equation}
\bar{\nu}_e+^{14}N\rightarrow e^++^{14}C
\end{equation}
where $\bar{\nu}_e$ is an electron anti-neutrino and $e^+$ is an antielectron, a positron. If the $\bar{\nu}_e$ 
spin (1/2, in units of $\hbar$, Planck's constant divided by 2$\pi$) is antiparallel to the 
$^{14}$N (spin 1), then the total spin of 1/2 on the 
left-hand side of the equation will equal the sum of the spin of $^{14}$C (spin 0) and the positron (spin 1/2) 
on the right-hand side. 
However, if the $\bar{\nu}_e$ spin and the $^{14}$N spins are aligned, then conservation of angular momentum will require one unit of angular momentum to come from either the $\bar{\nu}_e$ wave function or the positron wave function in order for the total angular momentum on the right-hand side 
to equal the 3/2 on the left-hand side. This is known from basic nuclear physics \citep{boyd08} to introduce roughly an order of 
magnitude smaller cross section for the latter case compared to the former, and is the origin of the effect predicted for the SNAAP 
model.

Detailed quantum 
molecular calculations have shown that the complex interactions of the molecules with the intense magnetic field of the nascent 
neutron star in a developing supernova or of the cooling neutron star following a supernova event and the electric field caused by 
the motion of the meteoroids through the magnetic field do produce an environment that is truly chiral 
\citep{barron86,barron08}. 
In this situation, the interactions of the $^{14}$N with the $\bar{\nu}_e$s are chirally selective, 
and will, at least in nearly every 
case, destroy more of the right-handed amino acids than the left-handed ones \citep{famiano18b}. 

The meteoroids that are processed by the anti-neutrinos can be as large as needed to survive the possibly intense fields of the stellar object they pass by or orbit. That isn't a particularly stringent assumption, since all that is needed is the magnetic field and the anti-neutrino flux, and there are several candidates that appear capable of satisfying those requirements: supernovae, cooling neutron stars, magnetars, Wolf-Rayet stars, and even ``silent supernovae;'' stars that are sufficiently massive that they collapse to black holes, develop strong magnetic fields, and emit the usual copious streams of neutrinos and anti-neutrinos while producing very few photons. 

Calculations were performed \citep{famiano18a,famiano18b} with the quantum molecular code \texttt{Gaussian} to examine several possible ways 
in which the $^{14}$N, coupled to the molecular chirality, could undergo chirality dependent destruction. 
This was done for twenty one 
amino acids. The motion of the meteoroids in the magnetic field of the central object is critical, as it induces an electric field 
from the cross product of the velocity with that magnetic field. The angle that the nuclear magnetization makes with the 
anti-neutrino spin is then chirally dependent. The cross section for destruction of 
the $^{14}$N by the anti-neutrinos, hence of the 
molecule, depends on that angle, producing the chirality dependent molecular destruction. 

The most promising scenario of the several studied \citep{famiano18a,famiano18b}
appears to result from the coupling 
of the molecular electric dipole moment to the electric field induced in the meteoroid by its motion. This produces transverse 
magnetization components that differ between the two molecular chiral states. These components exist even without the coupling to 
the electric dipole moment \citep{buckingham04,buckingham06}, but that coupling enhances the difference between 
the angles that the two chiral states make with the anti-neutrino spin, hence of the chirality selective destruction of the amino 
acids \citep{famiano18a}. From the magnitude of these effects, one can determine the ees that might be expected for amino 
acids from the SNAAP model.

In principle, electron neutrinos could drive the $^{14}$N to $^{14}$O, but the threshold energy is higher for this reaction. Since the cross section for neutrino capture processes is proportional to the square of the energy above threshold \citep{boyd08} this reaction has a smaller effect on the enantiomerism that results from the combined flux from anti-neutrinos and neutrinos.
\section{Results}
\label{results}
\paragraph{Can the SNAAP model produce ees in the amino acids?}
At present the quantum molecular calculations have assumed that the meteoroids pass by the central object, if it is a supernova or 
cooling neutron star, at mid-plane and normal to the axis that connects the poles. But the resulting $ee$s, as high as one percent, 
with the amino acid isovaline in an aqueous environment, as has been suggested in recent meteoritic analyses
\citep{herd11}, 
are particularly noteworthy in that they are in the ballpark of what is observed in the meteorites. However, if more sophisticated 
calculations fail to increase the predicted ees over the one percent level, some autocatalysis will be necessary for the SNAAP 
model to explain the meteoritic $ee$s.
\paragraph{Can the SNAAP model produce sufficiently large ees that some autocatalysis can boost them to the levels observed in the meteoroids?}
The required level of any ee producing mechanism might be relaxed if autocatalysis 
\citep{frank53,kondepudi85,goldanskii89}
can prevail in outer space. The experiments that have demonstrated autocatalysis 
\citep{soai95,soai02,breslow06,klussman06,soai14}
have been performed in laboratory settings. Although the 
minimum ee required for that to take effect is not known, it can be safely assumed to be less than the roughly one percent level in 
the experiments in which it has been demonstrated. Since the SNAAP model appears capable of producing ees at roughly that level, 
the required ee should not be a problem at all, unless autocatalysis is more restrictive in the cold confines of outer space than 
it is on Earth. Of course, that is a possibility. Thus experiments to determine the temperature dependence of autocatalysis would 
be very useful.
\paragraph{Can the SNAAP model predict that some of the carbonaceous chondrite meteorites that get to Earth will have nonzero ees?}
\label{meteoroids}
In order for any model to explain how some of the carbonaceous chondrite meteorites end up having ees, the model must either have a 
well-defined local source that can produce ees, or it must explain how it can process the space debris in some larger region of 
space.

a) One possibility might be thought to be the processing of the planets around a single massive star as it becomes a supernova. KEPLER \citep{borucki16}
has now detected planets around many stars. Thus it might be safe to argue that most, or at least many, stars do have planets 
associated with them. The inner ones will be completely processed by the anti-neutrinos, since nearly all of them will pass through 
any object, even a planet, as the star becomes a supernova. When the shock wave from the explosion hits the inner planets a few 
hours later, material will undoubtedly be spalled off, creating meteoroids. However, this model has a fundamental problem for the 
SNAAP model (and others) in that the magnetic field from the nascent neutron star extends to about 1 A.U., whereas the star, when 
it moves into its Red Giant phase will extend to about that same distance. Thus any meteoroids or planets that had any amino acids 
prior to the Red Giant phase would most likely have them destroyed when the star expanded. Although supernovae may be a major 
source of the galaxy's space debris, the amino acids in the resulting meteoroids would most likely have tiny enantiomeric excesses. 

b) Another possible scenario might result from a neutron star that is recoiling, after it has been produced in a supernova, 
typically at 1000 km/s or less, through the space debris of the galaxy for the 10$^5$ years it would be expected to continue to 
emit anti-neutrinos, processing each nearby floating planet \citep{sumi11} or piece of space rock as it goes. We investigated this 
scenario, but found that, even with generous estimates of the supernova frequency and the energies of the anti-neutrinos emitted by 
the cooling neutron star (they may be thermal, as described by \cite{bahcall65}, but may also have considerably higher 
energies from the nuclear processes taking place in the cooling star, as noted by 
\cite{fuller91,schatz14,misch16} and \cite{patton17}), the volume of the space that could be processed by all the neutron stars 
produced since the Big Bang was more than 10 orders of magnitude less than the volume of the galaxy. Furthermore, the space rocks 
so processed would be widely distributed, so would not be likely to populate a restricted region of space. 

c) A third possibility might be a Wolf-Rayet star. When the star became a supernova any amino acids that resided within a passing 
meteoroid or in the material surrounding the star within an A.U. of the star would be processed by the magnetic field and 
anti-neutrinos emitted. This does seem to be a plausible scenario for creating enantiomerism in the amino acids, although the 
trajectory of the passing meteoroid couldn't be too close to the hot star or too far from it to experience its magnetic field when 
it exploded. And dust grains within the surrounding cloud would have to have been in a sufficiently cool region for amino acids to form.

d) Perhaps a more likely scenario is one in which a massive star exists as part of a close binary system in which the partner is a 
neutron star. In such a system, the neutron star gradually siphons off the outer layer of the massive star, producing a 
star that will ultimately become a Type Ib/c supernova, and creating an accretion disk around the neutron star \citep{wolsczan08}. 
The 
disk apparently ranges from close to, but slightly beyond, the disk surface of the neutron star \citep[see, e.g.,][]{ludlam17}
to 
beyond 10$^5$ km \citep[see, e.g.,][]{pringle82}. 
The material would all be well inside the volume in which the combined magnetic field from 
the neutron star and from the supernova when it occurred would be sufficient to provide the necessary magnetic orientation, and 
close enough to the massive star to be subject to a robust anti-neutrino flux, when it exploded. This scenario introduces a complex 
set of possibilities. Any planets that were in orbits around the massive star would lose some of their gravitational attraction to 
that star as its mass was transferred to the neutron star, so that those in outer orbits might assume new, possibly highly 
elongated, orbits around the binary-star system \citep[see, e.g.,][]{jain17}, or might undergo a hyperbolic trajectory pass-by of 
the neutron star. In either scenario, the planet might be shredded by the strong gravitational field gradient, or as it passed 
through the accretion disk, so the result might produce the mass of the planet in meteoroids. 

The accretion disk itself is thought to be a nursery for dust grains, meteoroids, and even planets \citep{lithwick09}, and the 
temperature falloff with radius in the disk, thought to be $r^{-3/4}$ for large enough distance from the neutron 
star \citep[see, e.g.,][]{mineshige94}, 
would eventually provide a sufficiently low temperature environment in the outer regions of the disk that 
racemic amino acids could form, awaiting the anti-neutrinos from the exploding supernova to create some enantiomerism. The 
anti-neutrinos emitted by the cooling neutron star might become thermal soon after the neutron star is created so, except for those 
far out on the high energy tail of the distribution, their energy would be insufficient to cause the conversion of $^{14}$N to $^{14}$C. 
However, as noted above, nuclear processes might modify that conclusion. But when the massive star companion became a supernova, 
the matter in the accretion disk would all be well within the range of the neutrinos emitted from the supernova, which would 
process any amino acids that had developed in the accretion disk. Furthermore, the intense emissions from the X-ray binary and the 
shock wave from the supernova would surely cause sufficient disruption of at least some of the material in the disk to propel it 
beyond the gravitational well of the two stars.

What would happen to the binary system that had now become two neutron stars? Recent gravitational wave and space borne gamma ray detectors \citep{abbott17,goldstein17,savchenko17}
have shown that  neutron star mergers can produce a huge abundance of neutron-rich material, 
and presumably enough of an accompanying shock wave to create a new stellar system from the 
material ejected from what was originally two massive stars. This system may be capable of 
creating its own new stellar system, complete with r-process nuclides and enantiomeric amino acids. 

e) A recent study \citep{schatz14} of the crust in a neutron star deserves special note. It suggests that the nuclei that are contained in the matter that is 
accreted from the companion star into the neutron star accretion disk, and subsequently onto the surface of the neutron star, would 
be absorbed into the surface region of the star. They would encounter the essentially neutron pure matter ultimately to 
a depth of about 150 
meters, and would be driven to the neutron drip line by successive beta-decays and electron captures. The processes that would 
occur in one of the shells of the star would be
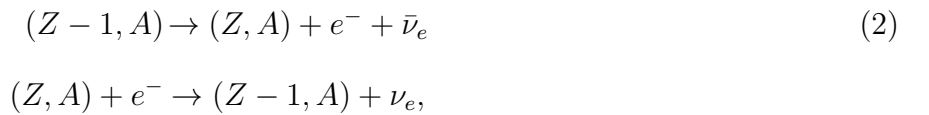
\begin{eqnarray}
(Z-1, A)&\rightarrow(Z,A) + e^- + \bar{\nu}_e\\\nonumber
(Z,A) + e^-&\rightarrow (Z-1, A) +  \nu_e,
\end{eqnarray}
where (Z,A) is a nucleus with proton number Z and nucleon number A. The star is cooled by the emission of the neutrinos, $\nu_e$, and anti-neutrinos, $\bar{\nu}_e$. As the nuclides are pushed more deeply into the neutron rich region below the crust, they become increasingly neutron rich until they reach the neutron drip line. The result could be 
a so-called URCA process 
\citep{gamow41} that would emit electron neutrinos and anti-neutrinos.

The anti-neutrino end point energies would be expected to achieve several MeV for some of the neutron-rich nuclides created.  
While the 
intensity of the resulting anti-neutrinos would not be as high as those emitted when the supernova explodes, they would be high 
enough in energy to process any amino acids that had been produced. Furthermore, they could continue to be processed 
for years, creating an additional
opportunity to process any amino acids created in the accretion disk around a neutron star from the electron anti-neutrinos emitted.

Thus this scenario might enhance the enantiomerism produced in the accretion disk in a binary system discussed in Section \ref{meteoroids}d. 
\\

Could this model populate the entire Galaxy with enantiomeric amino acids? That is very doubtful. WR stars and binary systems of 
the type discussed are not extremely rare, but neither do they occur frequently. However the meteoroids thrown out from the 
accretion disk of the binary system or the WR star would attain enough momentum from the Type Ib/c supernova to carry them to 
appreciable distances from the central system, and thus to populate a region that would ultimately be considerably larger than the 
Solar system. 

This would suggest that, although the potential for life would not be uniform throughout the Galaxy, there should be numerous 
pockets in which life might have been initiated as the enantiomeric amino acids were distributed around the binary star systems. 
Even though planets might lie in the Goldilocks zone, that is, within a temperature range that is neither too hot nor too cold for life to exist, 
they might not have amino acids that had received the necessary processing to make them enantiomeric. However, there might also be 
systems, specifically remnants of binary massive star systems, that would be strong candidates for life. Indeed, if the SNAAP model 
is the correct description of amino acid enantiomeric excess production, remnants of such systems should provide good places for 
astronomers to search for chiral amino acids. 
\paragraph{Can the SNAAP model produce meteorites that can make it to Earth's surface?}
Since the anti-neutrinos will have processed the entire meteoroid, no matter how large it was, the $ee$s established would prevail 
throughout its body. Thus, assuming that some of the resulting meteoroids would be large enough to suffer some ablation in passing 
through the Earth's atmosphere, whatever portion remained would carry the $ee$s it achieved prior to entering Earth's atmosphere. 
Dust grains would not be so fortunate; they would be likely to burn up before reaching the surface of the Earth.
\section{Some Conclusions}
\label{conclusion}
Several effects that are beyond the scope of the current paper will be dealt with in future studies. These include more 
calculations of quantum molecular chemistry and inclusion of time changing magnetic fields. Although we cannot be sure how these 
will affect the $ee$s, those calculated in the simplified model assumed in \cite{famiano18a}
were approaching the levels found in 
the meteorites. Thus the SNAAP model may require little, if any, outer space autocatalysis to produce the few percent $ee$s seen in 
meteorites. 

Perhaps the most troublesome aspect of the SNAAP model is that its $ee$ predictions are, at this stage, completely theoretical. 
Although the calculated $ee$s are the result of state of the art quantum molecular codes, it would be helpful to that model if some 
experiments could be performed to demonstrate the viability of at least some of its predictions. Experiments do appear to be 
feasible, and are under consideration. None the less, there do seem to be several plausible sites that apparently could produce the 
necessary magnetic fields and anti-neutrino fluxes to convert the amino acids produced in the outer reaches of the accretion disk 
from racemic to the slightly enantiomeric values found in some of the meteorites that made it to the surface of the earth. 

The predictions from this model are compelling. The enantiomeric levels achieved are approaching the levels seen in the 
meteorites, even without autocatalysis. And the possibility of a massive-star-neutron-star binary system being able to produce 
pockets of enantioermic amino acids suggests that this might well be the origin of the molecules found in the meteorites, and 
perhaps even of those required to initiate life one early Earth. It might behoove astronomers, when they are able to detect amino 
acids in space, to direct their efforts to determine enantiomerism toward the regions around close massive-star-neutron-star 
binaries.

We note that another scenario for producing enantiomeric amino acids in outer space, that of 
\cite{barron84} and \cite{rikken97}
would be facilitated by the sites we discuss above. This magneto-chiral dichroism model utilizes the light and a parallel 
magnetic field from a supernova to process the previously created amino acids. A single supernova would not suffice for the reason 
it doesn't suffice for the SNAAP model: the size of the Red Giant produced as one stage of the stellar evolution would extend 
beyond the region that could be served by the magnetic field of the nascent neutron star, a requirement of that model. However, the 
Wolf-Rayet star or a binary system obviates that issue, thus providing several sites described above for that model also.

\acknowledgments
The authors thank L. Nittler, F. Thielemann, H. Schatz, and L.D. Barron for helpful comments and suggestions. 
MAF's work was supported by the National Astronomical Observatory of Japan, and by a WMU Faculty Research and Creative
Activities Award (FRACAA). TK was supported partially by Grants-in-Aid for Scientific Research of JSPS (15H03665,
17K05459).


\software{Gaussian \citep{g16}}





\begin{thebibliography}{}
	\expandafter\ifx\csname natexlab\endcsname\relax\def\natexlab#1{#1}\fi
	\providecommand{\url}[1]{\href{#1}{#1}}
	
	\bibitem[{{Abbott} {et~al.}(2017){Abbott}, {Abbott}, {Abbott}, {Acernese},
		{Ackley}, {Adams}, {Adams}, {Addesso}, {Adhikari}, {Adya}, \&
		et~al.}]{abbott17}
	{Abbott}, B.~P., {Abbott}, R., {Abbott}, T.~D., {et~al.} 2017, \nat, 551, 85
	
	\bibitem[{{Bada} {et~al.}(1983){Bada}, {Cronin}, {Ho}, {Kvenvolden}, {Lawless},
		{Miller}, {Oro}, \& {Steinberg}}]{bada83}
	{Bada}, J.~L., {Cronin}, J.~R., {Ho}, M.-S., {et~al.} 1983, Nature, 301, 494
	
	\bibitem[{{Bahcall} \& {Wolf}(1965)}]{bahcall65}
	{Bahcall}, J.~N., \& {Wolf}, R.~A. 1965, Physical Review, 140, 1452
	
	\bibitem[{{Bailey} {et~al.}(1998){Bailey}, {Chrysostomou}, {Hough}, {Gledhill},
		{McCall}, {Clark}, {Menard}, \& {Tamura}}]{bailey98}
	{Bailey}, J., {Chrysostomou}, A., {Hough}, J.~H., {et~al.} 1998, Science, 281,
	672
	
	\bibitem[{Barron \& Vrbancich(1984)}]{barron84}
	Barron, L., \& Vrbancich, J. 1984, Molecular Physics, 51, 715.
	\newblock \url{https://doi.org/10.1080/00268978400100481}
	
	\bibitem[{Barron(1986)}]{barron86}
	Barron, L.~D. 1986, Journal of the American Chemical Society, 108, 5539.
	\newblock \url{http://dx.doi.org/10.1021/ja00278a029}
	
	\bibitem[{{Barron}(2008)}]{barron08}
	{Barron}, L.~D. 2008, Space Sciences Series, 135, 187
	
	\bibitem[{{Bonner}(1991)}]{bonner91}
	{Bonner}, W.~A. 1991, Origins of Life and Evolution of the Biosphere, 21, 59
	
	\bibitem[{{Borucki}(2016)}]{borucki16}
	{Borucki}, W.~J. 2016, Reports on Progress in Physics, 79, 036901
	
	\bibitem[{Boyd(2008)}]{boyd08}
	Boyd, R.~N. 2008, An Introduction to Nuclear Astrophysics (The University of
	Chicago Press (Chicago)), 126 -- 132
	
	\bibitem[{Boyd(2012)}]{boyd12}
	---. 2012, Stardust, Supernovae and the Molecules of Life (Springer-Verlag New
	York)
	
	\bibitem[{{Boyd} {et~al.}(2010){Boyd}, {Kajino}, \& {Onaka}}]{boyd10}
	{Boyd}, R.~N., {Kajino}, T., \& {Onaka}, T. 2010, Astrobiology, 10, 561
	
	\bibitem[{Boyd {et~al.}(2011)Boyd, Kajino, \& Onaka}]{boyd11}
	Boyd, R.~N., Kajino, T., \& Onaka, T. 2011, International Journal of Molecular
	Sciences, 12, 3432 .
	\newblock \url{http://www.mdpi.com/1422-0067/12/6/3432}
	
	\bibitem[{Breslow \& Levine(2006)}]{breslow06}
	Breslow, R., \& Levine, M.~S. 2006, Proceedings of the National Academy of
	Sciences, 103, 12979 .
	\newblock \url{http://www.pnas.org/content/103/35/12979.abstract}
	
	\bibitem[{Buckingham(2004)}]{buckingham04}
	Buckingham, A. 2004, Chemical Physics Letters, 398, 1 .
	\newblock \url{//www.sciencedirect.com/science/article/pii/S0009261404012308}
	
	\bibitem[{Buckingham \& Fischer(2006)}]{buckingham06}
	Buckingham, A., \& Fischer, P. 2006, Chemical Physics, 324, 111 .
	\newblock
	\url{http://www.sciencedirect.com/science/article/pii/S0301010405004970}
	
	\bibitem[{Cronin {et~al.}(1998)Cronin, Pizzarello, \& Cruikshank}]{cronin98}
	Cronin, J., Pizzarello, S., \& Cruikshank, D. 1998, in Meteorites and the Early
	Solar System, ed. J.~Kerridge \& M.~Matthews (Tucson: University of Arizona
	Press), 819 -- 857
	
	\bibitem[{{Cronin} \& {Pizzarello}(1997)}]{cronin97}
	{Cronin}, J.~R., \& {Pizzarello}, S. 1997, Science, 275, 951
	
	\bibitem[{{de Marcellus} {et~al.}(2011){de Marcellus}, {Meinert}, {Nuevo},
		{Filippi}, {Danger}, {Deboffle}, {Nahon}, {Le Sergeant d'Hendecourt}, \&
		{Meierhenrich}}]{demarcellus11}
	{de Marcellus}, P., {Meinert}, C., {Nuevo}, M., {et~al.} 2011, Astrophysical
	Journal Letters, 727, L27
	
	\bibitem[{Famiano {et~al.}(2018{\natexlab{a}})Famiano, Boyd, Kajino, \&
		Onaka}]{famiano18a}
	Famiano, M., Boyd, R., Kajino, T., \& Onaka, T. 2018{\natexlab{a}},
	Astrobiology, 18, XXX , (in press, published online).
	\newblock \url{https://doi.org/10.1089/ast.2017.1686}
	
	\bibitem[{Famiano {et~al.}(2014)Famiano, Boyd, Kajino, Onaka, Koehler, \&
		Hulbert}]{famiano14}
	Famiano, M., Boyd, R., Kajino, T., {et~al.} 2014, Symmetry, 6, 909 .
	\newblock \url{http://www.mdpi.com/2073-8994/6/4/909}
	
	\bibitem[{Famiano \& Boyd(2016)}]{famiano16}
	Famiano, M.~A., \& Boyd, R.~N. 2016, in Handbook of Supernovae, ed. A.~W.
	Alsabti \& P.~Murdin (Springer International Publishing), 1 -- 17.
	\newblock \url{http://dx.doi.org/10.1007/978-3-319-20794-0_20-1}
	
	\bibitem[{Famiano {et~al.}(2018{\natexlab{b}})Famiano, Boyd, Kajino, Onaka, \&
		Mo}]{famiano18b}
	Famiano, M.~A., Boyd, R.~N., Kajino, T., Onaka, T., \& Mo, Y.
	2018{\natexlab{b}}, Scientific Reports, XXX, XXX, submitted
	
	\bibitem[{Flack(2009)}]{flack09}
	Flack, H.~D. 2009, Acta Crystallographica Section A, 65, 371.
	\newblock \url{https://doi.org/10.1107/S0108767309024088}
	
	\bibitem[{Flores {et~al.}(1977)Flores, Bonner, \& Massey}]{flores77}
	Flores, J.~J., Bonner, W.~A., \& Massey, G.~A. 1977, Journal of the American
	Chemical Society, 99, 3622.
	\newblock \url{http://dx.doi.org/10.1021/ja00453a018}
	
	\bibitem[{Frank(1953)}]{frank53}
	Frank, F. 1953, Biochimica et Biophysica Acta, 11, 459 .
	\newblock
	\url{http://www.sciencedirect.com/science/article/pii/0006300253900821}
	
	\bibitem[{Frisch {et~al.}(2016)Frisch, Trucks, Schlegel, Scuseria, Robb,
		Cheeseman, Scalmani, Barone, Petersson, Nakatsuji, Li, Caricato, Marenich,
		Bloino, Janesko, Gomperts, Mennucci, Hratchian, Ortiz, Izmaylov, Sonnenberg,
		Williams-Young, Ding, Lipparini, Egidi, Goings, Peng, Petrone, Henderson,
		Ranasinghe, Zakrzewski, Gao, Rega, Zheng, Liang, Hada, Ehara, Toyota, Fukuda,
		Hasegawa, Ishida, Nakajima, Honda, Kitao, Nakai, Vreven, Throssell,
		Montgomery, Peralta, Ogliaro, Bearpark, Heyd, Brothers, Kudin, Staroverov,
		Keith, Kobayashi, Normand, Raghavachari, Rendell, Burant, Iyengar, Tomasi,
		Cossi, Millam, Klene, Adamo, Cammi, Ochterski, Martin, Morokuma, Farkas,
		Foresman, \& Fox}]{g16}
	Frisch, M.~J., Trucks, G.~W., Schlegel, H.~B., {et~al.} 2016, Gaussian˜16
	{R}evision {A}.03, , , gaussian Inc. Wallingford CT
	
	\bibitem[{{Fuller} \& {Meyer}(1991)}]{fuller91}
	{Fuller}, G.~M., \& {Meyer}, B.~S. 1991, \apj, 376, 701
	
	\bibitem[{{Gamow} \& {Schoenberg}(1941)}]{gamow41}
	{Gamow}, G., \& {Schoenberg}, M. 1941, Physical Review, 59, 539
	
	\bibitem[{{Glavin} \& {Dworkin}(2009)}]{glavin09}
	{Glavin}, D.~P., \& {Dworkin}, J.~P. 2009, Proceedings of the National Academy
	of Science, 106, 5487
	
	\bibitem[{{Goldanskii}(1989)}]{goldanskii89}
	{Goldanskii}, V.~I. 1989, Origins of Life and Evolution of the Biosphere, 19,
	269
	
	\bibitem[{{Goldstein} {et~al.}(2017){Goldstein}, {Veres}, {Burns}, {Briggs},
		{Hamburg}, {Kocevski}, {Wilson-Hodge}, {Preece}, {Poolakkil}, {Roberts},
		{Hui}, {Connaughton}, {Racusin}, {von Kienlin}, {Dal Canton}, {Christensen},
		{Littenberg}, {Siellez}, {Blackburn}, {Broida}, {Bissaldi}, {Cleveland},
		{Gibby}, {Giles}, {Kippen}, {McBreen}, {McEnery}, {Meegan}, {Paciesas}, \&
		{Stanbro}}]{goldstein17}
	{Goldstein}, A., {Veres}, P., {Burns}, E., {et~al.} 2017, \apjl, 848, L14
	
	\bibitem[{Guijarro \& Yus(2009)}]{guijarro09}
	Guijarro, A., \& Yus, M. 2009, The Origin of Chirality in the Molecules of Life
	(The Royal Society of Chemistry), P001--150, doi:10.1039/9781847558756.
	\newblock \url{http://dx.doi.org/10.1039/9781847558756}
	
	\bibitem[{{Herd} {et~al.}(2011){Herd}, {Blinova}, {Simkus}, {Huang}, {Tarozo},
		{Alexander}, {Gyngard}, {Nittler}, {Cody}, {Fogel}, {Kebukawa}, {Kilcoyne},
		{Hilts}, {Slater}, {Glavin}, {Dworkin}, {Callahan}, {Elsila}, {De Gregorio},
		\& {Stroud}}]{herd11}
	{Herd}, C.~D.~K., {Blinova}, A., {Simkus}, D.~N., {et~al.} 2011, Science, 332,
	1304
	
	\bibitem[{{Jain} {et~al.}(2017){Jain}, {Paul}, {Sharma}, {Jaleel}, \&
		{Dutta}}]{jain17}
	{Jain}, C., {Paul}, B., {Sharma}, R., {Jaleel}, A., \& {Dutta}, A. 2017,
	\mnras, 468, L118
	
	\bibitem[{Klussmann {et~al.}(2006)Klussmann, Iwamura, Mathew, David H.~Wells,
		Pandya, Armstrong, \& Blackmond}]{klussman06}
	Klussmann, M., Iwamura, H., Mathew, S.~P., {et~al.} 2006, Nature, 441, 621
	
	\bibitem[{Kondepudi \& Nelson(1985)}]{kondepudi85}
	Kondepudi, D.~K., \& Nelson, G.~W. 1985, Nature, 314, 438
	
	\bibitem[{{Kvenvolden} {et~al.}(1970){Kvenvolden}, {Lawless}, {Pering},
		{Peterson}, {Flores}, \& {Ponnamperuma}}]{kvenvolden70}
	{Kvenvolden}, K., {Lawless}, J., {Pering}, K., {et~al.} 1970, Nature, 228, 923
	
	\bibitem[{{Lithwick}(2009)}]{lithwick09}
	{Lithwick}, Y. 2009, \apj, 693, 85
	
	\bibitem[{{Ludlam} {et~al.}(2017){Ludlam}, {Miller}, {Bachetti}, {Barret},
		{Bostrom}, {Cackett}, {Degenaar}, {Di Salvo}, {Natalucci}, {Tomsick},
		{Paerels}, \& {Parker}}]{ludlam17}
	{Ludlam}, R.~M., {Miller}, J.~M., {Bachetti}, M., {et~al.} 2017, \apj, 836, 140
	
	\bibitem[{Mason(1984)}]{mason84}
	Mason, S.~F. 1984, Nature, 311, 19
	
	\bibitem[{Meierhenrich(2008)}]{meierhenrich08}
	Meierhenrich, U. 2008, Amino Acids and the Asymmetry of Life: Caught in the Act
	of Formation (Springer-Verlag Verlin Heidelberg)
	
	\bibitem[{Meierhenrich {et~al.}(2010)Meierhenrich, Filippi, Meinert,
		Bredehöft, Takahashi, Nahon, Jones, \& Hoffmann}]{meierhenrich10}
	Meierhenrich, U.~J., Filippi, J.-J., Meinert, C., {et~al.} 2010, Angewandte
	Chemie International Edition, 49, 7799 .
	\newblock \url{http://dx.doi.org/10.1002/anie.201003877}
	
	\bibitem[{Meierhenrich {et~al.}(2005)Meierhenrich, Nahon, Alcaraz, Bredehöft,
		Hoffmann, Barbier, \& Brack}]{meierhenrich05}
	Meierhenrich, U.~J., Nahon, L., Alcaraz, C., {et~al.} 2005, Angewandte Chemie
	International Edition, 44, 5630.
	\newblock \url{http://dx.doi.org/10.1002/anie.200501311}
	
	\bibitem[{Meinert {et~al.}(2014)Meinert, Hoffmann, Cassam-Chenaï, Evans, Giri,
		Nahon, \& Meierhenrich}]{meinert14}
	Meinert, C., Hoffmann, S.~V., Cassam-Chenaï, P., {et~al.} 2014, Angewandte
	Chemie International Edition, 53, 210.
	\newblock \url{http://dx.doi.org/10.1002/anie.201307855}
	
	\bibitem[{Meinert {et~al.}(2010)Meinert, Filippi, Nahon, Hoffmann,
		D’Hendecourt, De~Marcellus, Bredehöft, Thiemann, \&
		Meierhenrich}]{meinert10}
	Meinert, C., Filippi, J.-J., Nahon, L., {et~al.} 2010, Symmetry, 2, 1055 .
	\newblock \url{http://www.mdpi.com/2073-8994/2/2/1055}
	
	\bibitem[{Meinert {et~al.}(2012)Meinert, Bredehöft, Filippi, Baraud, Nahon,
		Wien, Jones, Hoffmann, \& Meierhenrich}]{meinert12}
	Meinert, C., Bredehöft, J.~H., Filippi, J.-J., {et~al.} 2012, Angewandte
	Chemie International Edition, 51, 4484.
	\newblock \url{http://dx.doi.org/10.1002/anie.201108997}
	
	\bibitem[{{Miller}(1953)}]{miller53}
	{Miller}, S.~L. 1953, Science, 117, 528
	
	\bibitem[{{Miller} \& {Urey}(1959)}]{miller59}
	{Miller}, S.~L., \& {Urey}, H.~C. 1959, Science, 130, 245
	
	\bibitem[{{Mineshige} {et~al.}(1994){Mineshige}, {Honma}, {Hirano}, {Kitamoto},
		{Yamada}, \& {Fukue}}]{mineshige94}
	{Mineshige}, S., {Honma}, F., {Hirano}, A., {et~al.} 1994, in NATO Advanced
	Science Institutes (ASI) Series C, Vol. 417, NATO Advanced Science Institutes
	(ASI) Series C, ed. W.~J. {Duschl}, J.~{Frank}, F.~{Meyer},
	E.~{Meyer-Hofmeister}, \& W.~M. {Tscharnuter}, 187
	
	\bibitem[{{Misch} \& {Fuller}(2016)}]{misch16}
	{Misch}, G.~W., \& {Fuller}, G.~M. 2016, \prc, 94, 055808
	
	\bibitem[{{Norden}(1977)}]{norden77}
	{Norden}, B. 1977, Nature, 266, 567
	
	\bibitem[{Pasteur(1848)}]{pasteur48}
	Pasteur, L. 1848, Comptes Rendus Hebdomadaires des S\'eances de l'Acad\'emie
	des Sciences, 26, 535
	
	\bibitem[{{Patton} {et~al.}(2017){Patton}, {Lunardini}, {Farmer}, \&
		{Timmes}}]{patton17}
	{Patton}, K.~M., {Lunardini}, C., {Farmer}, R.~J., \& {Timmes}, F.~X. 2017,
	\apj, 851, 6
	
	\bibitem[{{Pringle}(1982)}]{pringle82}
	{Pringle}, J.~E. 1982, in Accreting Neutron Stars, ed. W.~{Brinkmann} \&
	J.~{Truemper}
	
	\bibitem[{Rikken \& Raupach(1997)}]{rikken97}
	Rikken, G., \& Raupach, E. 1997, Nature, 390, 493
	
	\bibitem[{{Rubenstein} {et~al.}(1983){Rubenstein}, {Bonner}, {Noyes}, \&
		{Brown}}]{rubenstein83}
	{Rubenstein}, E., {Bonner}, W.~A., {Noyes}, H.~P., \& {Brown}, G.~S. 1983,
	\nat, 306, 118
	
	\bibitem[{{Savchenko} {et~al.}(2017){Savchenko}, {Ferrigno}, {Kuulkers},
		{Bazzano}, {Bozzo}, {Brandt}, {Chenevez}, {Courvoisier}, {Diehl}, {Domingo},
		{Hanlon}, {Jourdain}, {von Kienlin}, {Laurent}, {Lebrun}, {Lutovinov},
		{Martin-Carrillo}, {Mereghetti}, {Natalucci}, {Rodi}, {Roques}, {Sunyaev}, \&
		{Ubertini}}]{savchenko17}
	{Savchenko}, V., {Ferrigno}, C., {Kuulkers}, E., {et~al.} 2017, \apjl, 848, L15
	
	\bibitem[{{Schatz} {et~al.}(2014){Schatz}, {Gupta}, {M{\"o}ller}, {Beard},
		{Brown}, {Deibel}, {Gasques}, {Hix}, {Keek}, {Lau}, {Steiner}, \&
		{Wiescher}}]{schatz14}
	{Schatz}, H., {Gupta}, S., {M{\"o}ller}, P., {et~al.} 2014, \nat, 505, 62
	
	\bibitem[{Soai {et~al.}(2014)Soai, Kawasaki, \& Matsumoto}]{soai14}
	Soai, K., Kawasaki, T., \& Matsumoto, A. 2014, The Chemical Record, 14, 70 .
	\newblock \url{http://dx.doi.org/10.1002/tcr.201300028}
	
	\bibitem[{Soai \& Sato(2002)}]{soai02}
	Soai, K., \& Sato, I. 2002, Chirality, 14, 548 .
	\newblock \url{http://dx.doi.org/10.1002/chir.10081}
	
	\bibitem[{Soai {et~al.}(1995)Soai, Shibata, Morioka, \& Choji}]{soai95}
	Soai, K., Shibata, T., Morioka, H., \& Choji, K. 1995, Nature, 378, 767 .
	\newblock
	\url{http://http://www.nature.com/nature/journal/v378/n6559/abs/378767a0.html}
	
	\bibitem[{{Sumi} {et~al.}(2011){Sumi}, {Kamiya}, {Bennett}, {Bond}, {Abe},
		{Botzler}, {Fukui}, {Furusawa}, {Hearnshaw}, {Itow}, {Kilmartin}, {Korpela},
		{Lin}, {Ling}, {Masuda}, {Matsubara}, {Miyake}, {Motomura}, {Muraki},
		{Nagaya}, {Nakamura}, {Ohnishi}, {Okumura}, {Perrott}, {Rattenbury}, {Saito},
		{Sako}, {Sullivan}, {Sweatman}, {Tristram}, {Udalski}, {Szyma{\'n}ski},
		{Kubiak}, {Pietrzy{\'n}ski}, {Poleski}, {Soszy{\'n}ski}, {Wyrzykowski},
		{Ulaczyk}, \& {Microlensing Observations in Astrophysics (MOA)
			Collaboration}}]{sumi11}
	{Sumi}, T., {Kamiya}, K., {Bennett}, D.~P., {et~al.} 2011, \nat, 473, 349
	
	\bibitem[{Takahashi {et~al.}(2009)Takahashi, Shinojima, Seyama, Ueno, Kaneko,
		Kobayashi, Mita, Adachi, Hosaka, \& Katoh}]{takahashi09}
	Takahashi, J.-i., Shinojima, H., Seyama, M., {et~al.} 2009, International
	Journal of Molecular Sciences, 10, 3044 .
	\newblock \url{http://www.mdpi.com/1422-0067/10/7/3044}
	
	\bibitem[{{Takano} {et~al.}(2007){Takano}, {Takahashi}, {Kaneko}, {Marumo}, \&
		{Kobayashi}}]{takano07}
	{Takano}, Y., {Takahashi}, J.-i., {Kaneko}, T., {Marumo}, K., \& {Kobayashi},
	K. 2007, Earth and Planetary Science Letters, 254, 106
	
	\bibitem[{{Wolszczan}(2008)}]{wolsczan08}
	{Wolszczan}, A. 2008, Physica Scripta Volume T, 130, 014005
	
\end{thebibliography}


\end{document}